\documentclass{phb-proc4-auth}

\usepackage{graphicx}
\usepackage{amssymb}
\usepackage{amsfonts}
\usepackage{amsmath}
\usepackage{amscd}
\usepackage{cite}

\def\beq{\begin{equation}}
\def\eeq{\end{equation}}
\def\bea{\begin{eqnarray}}
\def\eea{\end{eqnarray}}

\def\beann{\begin{eqnarray*}}
\def\eeann{\end{eqnarray*}}

  \let\g=\gamma 
   
\let\eps=\epsilon
  \let\la=\lambda 
 \let\x=\xi \let\p=\pi  \let\s=\sigma
\let\om=\omega 
\let\ph=\varphi   
  
  \let\D=\Delta

\def\epp{\, .}
\def\epc{\, ,}

\def\tst#1{{\textstyle #1}}

\def\2{\frac{1}{2}} \def\4{\frac{1}{4}}

\def\6{\partial}

\def\+{\dagger}

\def\<{\langle} \def\>{\rangle}

\def\fa{\mathfrak{a}}

\begin{document}
\begin{frontmatter}


\journal{SCES '04}


\title{Emptiness formation probability at finite temperature for the
isotropic Heisenberg chain}

%
%
%
%
%
%

\author{Frank G\"ohmann,\corauthref{1}}
\author{Andreas Kl\"umper} and
\author{Alexander Seel}

%
 
\address{Bergische Universit\"at Wuppertal, 42097 Wuppertal, Germany}

%
%
%
%


%
%
%
%

\corauth[1]{Phone: ++49-202-439-2862, Fax: ++49-202-439-3860,\\
E-Mail: goehmann@physik.uni-wuppertal.de}


\begin{abstract}

We present an integral formula for a special correlation function
of the isotropic spin-1/2 antiferromagnetic Heisenberg chain. The
correlation function describes the probability for the occurrence of
a string of consecutive up-spins as a function of temperature,
magnetic field and length of the string.

\end{abstract}

%
%



\end{frontmatter}

%
%
%
%
%

In a recent work \cite{GKS04app} we derived (multiple) integral
representations for a generating function of the $\s^z$-$\s^z$
correlation functions of the spin-$\2$ XXZ chain,
\begin{equation} \label{xxzham}
     H = J \sum_{j=1}^L \Bigl( \s_{j-1}^x \s_j^x + \s_{j-1}^y \s_j^y
                               + \D (\s_{j-1}^z \s_j^z - 1) \Bigr) \epc
\end{equation}
at finite temperatures $T$ and finite values of an external,
longitudinal magnetic field $h$. Our formulae generalize the
zero-temperature results of \cite{KMST02a}.

In order to relate our work to experiments the integrals should be
calculated. Although this task may appear rather challenging at first
sight, the recent progress made in the calculation of related integrals
(see e.g. \cite{BoKo02,BKNS02,SSNT03}) and their asymptotic behaviour
\cite{KMST02d,KLNS03} at zero temperature lets us hope that we
eventually will be able to extract numbers from our formulae.

The integrals \cite{JMMN92,KIEU94,JiMi96} calculated in \cite{BoKo02,%
BKNS02,SSNT03} do not describe local correlation functions but density
matrix elements (termed `elementary blocks' in the mathematical
literature). One of these non-local correlation functions, called the
emptiness formation probability \cite{KIEU94}, is contained as a
special case in our formula obtained in \cite{GKS04app}. In
\cite{GKS04app} (see eqs.\ (125), (126)) we represented it as a
multiple integral with symmetric integrand of similar form as in
\cite{KMST02a}. Below we obtain another integral representation with
asymmetric integrand which, when further specialized to the isotropic
antiferromagnet ($J >~0$ and $\D = 1$ in (\ref{xxzham})), is the finite
temperature generalization of the formula obtained in \cite{KIEU94}
that served as a starting point for the explicit calculation of the
emptiness formation probability in \cite{BoKo02}. We therefore believe
that it may also be the appropriate starting point for proceeding
further to finite temperatures.

\section{Thermodynamics}
There are several alternative possibilities of representing the free
energy of the Heisenberg chain (\ref{xxzham}) as an integral over
auxiliary functions. The representation based on the quantum transfer
matrix \cite{Suzuki85,SuIn87} and derived in \cite{Kluemper92,%
Kluemper93} turns out to be most appropriate in the present context,
since the auxiliary functions involved also appear quite naturally in
the derivation of multiple integral representations of correlation
functions \cite{GKS04app}. Using the formulation of \cite{GKS04app} and
specializing to $J > 0$ and $\D = 1$ we can write the free energy per
lattice site in the thermodynamic limit as
\begin{equation} \label{freee}
     f (h,T) = - \frac{h}{2} - T \int_C \frac{d \om}{2 \p}\,
                 \frac{\ln (1 + \fa (\om))}
		      {\om (\om + {\rm i})} \epp
\end{equation}
The contour $C$ encloses the real axis at a distance $|\g| <~\2$.
\begin{figure}
    \centering
    \includegraphics{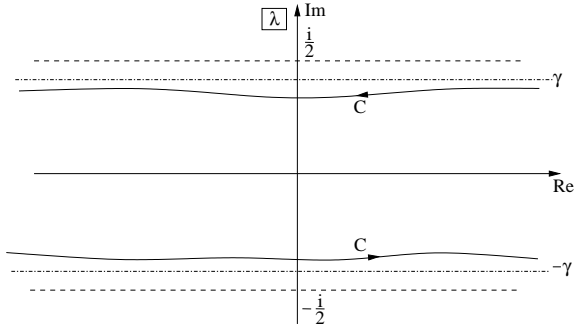}
    \caption{The canonical contour $C$.} 
\end{figure}  
The auxiliary function $\fa (\la)$ is the solution of the non-linear
integral equation
\begin{equation} \label{nlie}
     \ln \fa (\la) = - \frac{h}{T} + \frac{2J}{\la (\la + {\rm i})T}
                        - \int_C \frac{d \om}{\p}\,
			  \frac{\ln (1 + \fa (\om))}
			       {1 + (\la - \om)^2} \epp
\end{equation}

\section{Emptiness formation probability}
The emptiness formation probability \cite{KIEU94} is defined as the
thermal average
\begin{equation}
     P(m) = \bigl\< \tst{\prod_{j=1}^m} \:
            \bigl( \begin{smallmatrix} 1 & 0 \\ 0 & 0
	           \end{smallmatrix} \bigr)_{\mspace{-3.mu} j}
		   \bigr\>_T \epp
\end{equation}

Within the formalism developed in \cite{GKS04app} it is obtained as
the homogeneous limit
\begin{equation}
     P(m) = \lim_{\x_1, \dots, \x_m \rightarrow 0}
            P(m|\{\x_j\}_{j=1}^m)
\end{equation}
of a more abstract correlation function $P(m|\{\x_j\}_{j=1}^m)$.
Here we present a new integral representation for this function,
\begin{multline} \label{emptinh}
     P(m|\{\x_j\}_{j=1}^m) = \\ \Bigl[ \prod_{j=1}^m
        \int_C \frac{d \om_j}{2 \p (1 + \fa(\om_j))} \Bigr]
	\frac{\det G(\om_j, \x_k)}
	     {\prod_{1 \le j < k \le m} (\x_k - \x_j)} \cdot \\
        \cdot \frac{\prod_{j=1}^m
	   \bigl[ \prod_{k=1}^{j-1} (\x_k - \om_j + {\rm i}) \bigr]
	   \bigl[ \prod_{k = j + 1}^m (\x_k - \om_j) \bigr]}
	     {\prod_{1 \le j < k \le m} (\om_j - \om_k + {\rm i})} \epp
\end{multline}
As in our general formula in \cite{GKS04app} the function $G(\la,\x)$
appearing in the determinant on the right hand side has to be
calculated from a linear integral equation which in the isotropic case
($\D = 1$) reads
\begin{multline}
     G(\la,\x) = - \frac{1}{(\la - \x)(\la - \x - {\rm i})} \\ +
                   \int_C \frac{\d \om}{\p} \frac{1}{1 + (\la - \om)^2}
		          \frac{G(\om,\x)}{ 1 + \fa(\om)} \epp
\end{multline}
The formula (\ref{emptinh}) follows from equation (111) of
\cite{GKS04app} by (i) taking the limit $\ph \rightarrow - \infty$,
(ii) using appendix C of \cite{KMST02a}, and (iii) taking the
isotropic limit.

Taking the homogeneous limit of (\ref{emptinh}) is not difficult~%
\cite{KMT99b}. We obtain the expression
\begin{multline} \label{empth}
     P(m) = \Bigl[ \prod_{j=1}^m
        \int_C \frac{d \om_j}{2 \p (1 + \fa(\om_j))} \Bigr]
	   \cdot \\ \cdot
	\det \frac{\6^{(k-1)}_\x G(\om_j, \x)\bigr|_{\x = 0}}{(k-1)!}\:
        \frac{\prod_{j=1}^m (\om_j - {\rm i})^{j-1} \om_j^{m-j}}
	     {\prod_{1 \le j < k \le m} (\om_j - \om_k + {\rm i})} \epp
\end{multline}
This is the direct generalization to finite temperatures of the formula
obtained in \cite{JMMN92,KIEU94,KMT99b}.

\section{Reduction of the integrals}
We believe that either (\ref{emptinh}) or (\ref{empth}) or the
symmetrized form of (\ref{emptinh}) is a good starting point for
studying the possibility to reduce the multiple integrals to simpler
ones. That this may be possible in principle can be inferred from the
example $m = 2$, where, for $h = 0$ and in the isotropic limit, the
emptiness formation probability $P(2)$ can be expressed in terms of
the internal energy $\eps$,
\begin{equation}
     P(2) = \frac{1}{3} + \frac{\eps}{12 J} \epp
\end{equation}
Inserting (\ref{freee}) into the thermodynamic expression for the
internal energy $\eps = \6_{1/T} (f/T)$ we obtain
\begin{equation}
     P(2) = \frac{1}{3} - \frac{1}{12} \int_C \frac{d \om}{\p}
               \frac{1}{\om (\om - {\rm i})} \frac{G(\om,0)}{1 + \fa(\om)}
	       \epp
\end{equation}
Note that there is only a single integral on the right hand side.
Specializing (\ref{empth}) for $m = 2$, on the other hand, yields
\begin{multline} \label{empth2}
     P(2) = \\ \frac{1}{4 \p^2}
        \int_C \frac{d \om_1}{1 + \fa(\om_1)}
        \int_C \frac{d \om_2}{1 + \fa(\om_2)}
        \frac{\om_1 (\om_2 - {\rm i})}{\om_1 - \om_2 + {\rm i}} \cdot \\[1ex]
	\cdot \bigl(G(\om_1, 0) \6_\x G(\om_2, \x)
	   - G(\om_2, 0) \6_\x G(\om_1, \x)\bigr) \bigr|_{\x = 0}
\end{multline}
which involves two integrations.

\section{Discussion}
The integral representations (\ref{emptinh}), (\ref{empth}) are likely
candidates for extraxting for the first time explicit numbers for
finite temperature next-to-nearest neighbour correlation functions
of the isotropic Heisenberg chain and may serve as a model case to
initiate a systematic study of the more general integrals derived in
\cite{GKS04app}.

{\bf Acknowledgement.} The authors are indebted to H. E. Boos and
I. Peschel for valuable discussions.

%
%
%
%


%
%
%
%


\end{document}